\documentclass[notitlepage]{revtex4-1} 
\usepackage{graphicx}

\usepackage{hyperref} 					 
\usepackage{listings} 					 
\usepackage{color}    					 
\usepackage{natbib}   					 
\usepackage{subfig} 					 

\usepackage{booktabs}
\usepackage[ansinew]{inputenc}				     
\usepackage{psfrag}
\usepackage{pstricks}
\usepackage{amsfonts}
\usepackage{amsmath}
\usepackage{bm}
\usepackage{lipsum}
\usepackage{pgf}
\usepackage{tikz}
\usetikzlibrary{patterns}

\usepackage{algorithm}
\usepackage{algorithmic}

\usepackage{array}
\usepackage{booktabs}
\newcolumntype{C}[1]{>{\centering\arraybackslash}m{#1}}

\usepackage{filecontents}

\usepackage{amsmath}
\usepackage[misc,geometry]{ifsym} 
\begin{document}
\title{Topological Insulators by Topology Optimization}

\author{Rasmus E. Christiansen}
\email{Corresponding email: raelch@mek.dtu.dk}

\author{Fengwen Wang}

\author{Ole Sigmund}

\affiliation{Department of Mechanical Engineering, Solid Mechanics, Technical University of Denmark, Nils Koppels Allé, B.\ 404, DK-2800 Kgs. Lyngby, Denmark}


\begin{abstract}
\noindent An acoustic topological insulator (TI) is synthesized using topology optimization, a free material inverse design method. The TI appears spontaneously from the optimization process without imposing requirements on the existence of pseudo spin-1/2 states at the TI interface edge, or the Chern number of the topological phases. The resulting TI is passive; consisting of acoustically hard members placed in an air background and has an operational bandwidth of $\approx$12.5\% showing high transmission. Further analysis demonstrates confinement of more than 99\% of the total field intensity in the TI within at most six lattice constants from the TI interface. The proposed design hereby outperforms a reference from recent literature regarding energy transmission, field confinement and operational bandwidth. 

\end{abstract}

\maketitle

\begin{figure*}[!]
	\centering
	{
		\includegraphics[width=0.94\textwidth]{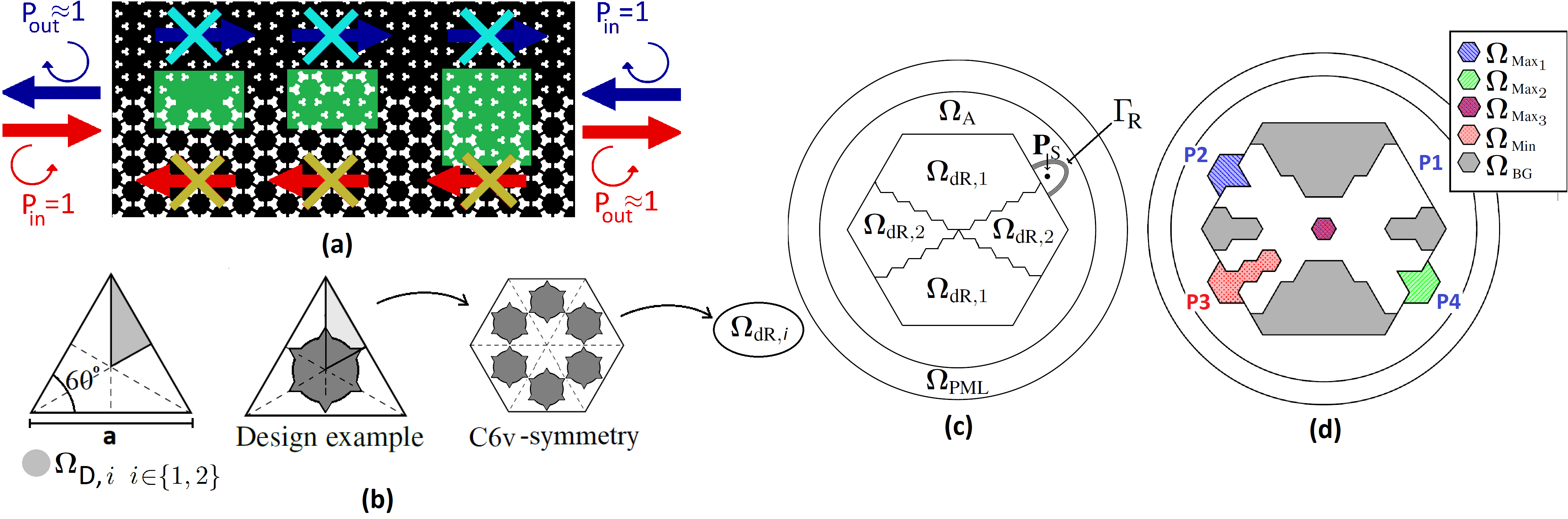}
		\caption{\textcolor{black}{\textbf{(a)} Illustration of backscattering protected, pseudo-spin dependent directional field propagation through a TI with [green] defects. \textbf{(b) left:} design cell showing the [grey] designable region and [dashed lines] mirror symmetry lines. \textbf{(b) middle and right:} Design example illustrating the symmetry. \textbf{(c)} Model domain $\bm{\Omega}_{\text{A}}$, PML layer $\bm{\Omega}_{\text{PML}}$ and domains $\bm{\Omega}_{\text{dR},1}$ and $\bm{\Omega}_{\text{dR},2}$ containing the two topological phases. Mono-polar source $\textbf{P}_\text{s}$ in the focal point of a reflector with surface $\Gamma_{\text{R}}$. \textbf{(d)} Domains for computing $\Phi_{\text{Total}}$ and $\Phi_{\text{BG}}$ and port numbering: P1-P4.} \ \label{FIG:MODEL_TARGET_DESIGN_DOMAIN} }
	}
\end{figure*}

The concept of the topological insulator (TI) stems from condensed matter physics and the quantum spin Hall effect (QSHE) \cite{THOULESS_1982,HALDANE_1988}. Following these seminal works, a growing effort has been dedicated to understanding and designing TIs \cite{HASAN_KANE_2010,QI_ZHANG_2011}, with works demonstrating the engineering of TIs within the fields of photonics \cite{KHANIKAEV_ET_AL_2012,RAGHU_2008,WANG_2008,WANG_2009,HAFEZI_2011,CHEN_2014,LU_2014,WU_2015}, solid mechanics \cite{SUSSTRUNK_2015,WANG_2015,MOUSAVI_2015} and acoustics \cite{NI_2015,YANG_ET_AL_2015,HE_2016,KHANIKAEV_2015,FLEURY_2016} alike. This surge in interest is partly fuelled by the incredible promise that TIs can provide backscattering protected, edge-state confined, one-way energy transport, robust under a class of structural defects. Such properties are obviously of broad interest, with numerous applications able to benefit from backscattering protected energy transport, a recent example being lasing \cite{ST-JEAN_ET_AL_2017}. Three fundamentally different systems for TIs are known: time-reversal breaking; time-reversal invariant; and Floquet topological systems, each providing different modes of operation \cite{KHANIKAEV_2017}. This letter considers the time-reversal invariant setting in acoustics, allowing for backscattering protected spin-dependent directional energy transport, robust towards defects, as illustrated in Fig. \ref{FIG:MODEL_TARGET_DESIGN_DOMAIN}(a). 

Acoustic systems intrinsically possess spin-0, thus no Kramers doublets exist, hindering the manifestation of the acoustic QSHE. This barrier can be overcome by constructing artificial acoustic spin-1/2 states, e.g. by creating circulating acoustic waves, actively \cite{YANG_ET_AL_2015} utilizing airflow or passively \cite{HE_2016} by engineering an accidental double Dirac cone through a change in the filling factor of cylindrical metallic rods in a honeycomb lattice. In \cite{HE_2016} a time-reversal invariant acoustic TI is engineered and shown to support topologically protected edge-states in a $\approx$1.5 kHz wide bulk-bandgap. The TI is demonstrated to largely suppress backscattering, with the measured transmission dropping at most 5 dB and to perform robustly under geometric defects, showing a maximum drop in transmission of 4 dB. 

As outlined above, significant effort has been invested in the design of TIs, leading to excellent results and new discoveries. The design procedures have, however, hitherto mainly been based on intuition, and the bottom-up approach of band-structure engineering. Such approaches do not consider the finite size of the physical structure, disregarding the coupling into and out of the TI. Further, approaches based on intuition are unlikely to lead to optimal designs, possibly leaving a large performance potential untapped. 

Inspired by the work in \cite{HE_2016} this letter proposes a fundamentally different, optimization based approach for the design of topological insulators. A top-down approach based on inverse design where the backscattering protected energy transport is targeted directly, with no explicit requirements on the underlying mechanisms or geometries. Hence, the approach does not impose requirements on the pre-existence of acoustic pseudo spin-1/2 edge-states; nor on the Chern numbers of the two involved topological phases; nor on band symmetry inversion in reciprocal space. These properties appear spontaneously during the design process. A TI designed using the proposed approach, is analysed and demonstrated to suppress backscattering from geometric defects while facilitating spin-dependent, directional energy transport and strong field confinement.

The proposed top-down design approach considers a carefully configured finite material slab; illuminated by an acoustic source; placed in a homogeneous background medium. It utilizes density based topology optimization \cite{BOOK_TOPOPT_BENDSOE} to solve the inverse design problem \textcolor{black}{starting from an initial guess provided by the user} and is inspired by work on designing meta-material slabs exhibiting negative refraction \cite{CHRISTIANSEN_SIGMUND_2016}. \textcolor{black}{It is noted that while the topology optimization method and the topological insulator share the word "topology" the two uses are not directly related. In topology optimization the word refers to the ultimate spatial design freedom that allows the algorithm to choose the structural topology which optimizes the objective function.} It is noteworthy that several recent works have demonstrated the benefit of using topology optimization in the design and optimization of exotic meta-materials and crystals, such as multifunctional optical meta-gratings \cite{SELL_ET_AL_2017}, elastic meta-materials with negative effective material parameters \cite{DONG_ET_AL_2016} and self-collimating phononic crystals \cite{PARK_ET_AL_2014}. \textcolor{black}{Further, two review papers \cite{JENSEN_SIGMUND_2011,MOLESKY_2018} for using inverse design in photonics show numerous successful uses of topology optimization.}

A sketch of the model domain serving as the design platform in this work, is shown in Fig. \ref{FIG:MODEL_TARGET_DESIGN_DOMAIN}(c). Here $\Omega_{\text{A}}$ denotes an air region surrounded by a perfectly matched layer \cite{Berenger_1994}, denoted $\Omega_{\text{PML}}$. A hexagonally shaped design domain is placed inside $\Omega_{\text{A}}$ and partitioned into the sub-domains $\Omega_{\text{dR},1}$ and $\Omega_{\text{dR},2}$ containing two different periodic structures (the topological phases). The slab is illuminated by a mono-polar point source, $\textbf{P}_{\text{S}}$, placed in the focal point of a perfectly reflecting parabolic reflector ($\textbf{n} \cdot \nabla \Psi(\textbf{r})  = 0 \ \forall \ \textbf{r} \in \Gamma_{\text{R}}$).

\textcolor{black}{The careful choice of the $180^{\text{o}}$ rotationally symmetric configuration of $\Omega_{\text{dR},1}$ and $\Omega_{\text{dR},2}$ is key to the proposed approach. This means that under ideal conditions any power flowing along the interface edge from P1 to P3 (see Fig. \ref{FIG:MODEL_TARGET_DESIGN_DOMAIN}(d)), will be indistinguishable from power flowing from P1 to the centre of the slab after which it reverses direction and flows back to P1. Hence, by minimizing the power flow from P1 to P3 one by extension minimizes back-scattering.}

The physics is modelled using a Helmholtz type equation,

\begin{equation} \label{EQN:HELMHOLTZ}
\nabla \cdot \left(  \text{C}_1 \frac{\omega_0 - \text{i} c \alpha}{\omega_0} \nabla \Psi \right) + \text{C}_2 \frac{(\omega_0-\text{i} c \alpha)^3}{\omega_0} \Psi = -\textbf{P}_{\text{S}}, 
\end{equation}

\noindent where $\text{C}_1(\textbf{r})$ and $\text{C}_2(\textbf{r})$ are material dependent parameters, $\text{i}$ the imaginary unit,  $\alpha(\textbf{r})$ an attenuation parameter, $\omega_0 = 2\pi f_0$ the free space angular frequency, $c$ the free space wave-speed, $\Psi(\textbf{r})$ the state field and $\textbf{r}$ the spatial position. For the acoustic case, $\Psi = p$ where $p(\textbf{r})$ is the sound pressure and $\lbrace \text{C}_1, \text{C}_2 \rbrace = \lbrace \frac{1}{\rho}, \frac{1}{\kappa} \rbrace$ where $\rho(\textbf{r})$ and $\kappa(\textbf{r})$ are the density and bulk modulus, respectively. Material parameters for air and aluminium are used \cite{DUHRING_JENSEN_SIGMUND_2008}. The impedance contrast between the two ensuring that vibrations exited in the solid are negligible, and thus \eqref{EQN:HELMHOLTZ} accurately captures the physics, as verified in \cite{Christiansen_Grande_Sigmund_2015,CHRISTIANSEN_SIGMUND_2016b}. 

The design problem is formulated as a continuous constrained optimization problem and solved using density based topology optimization. A spatial design field $\xi(\textbf{r}) \in [0,1] \ \forall \ \textbf{r} \in \ \Omega_{\text{dR},1} \bigcup \Omega_{\text{dR},2}$ is introduced to control the periodic material distributions in $\Omega_{\text{dR},1}$ and $\Omega_{\text{dR},2}$ by interpolating $\text{C}_1$ and $\text{C}_2$ between the material parameters as,

\begin{equation} \label{EQN:INTERPOLATION}
\text{C}_i^{-1}(\textbf{r}) = \text{C}_{i_{\text{air}}}^{-1} + \xi(\textbf{r})^6 \left(\text{C}_{i_{\text{aluminium}}}^{-1}-\text{C}_{i_{\text{air}}}^{-1}\right), i \in \lbrace 1, 2 \rbrace.
\end{equation}

\textcolor{black}{Figure \ref{FIG:MODEL_TARGET_DESIGN_DOMAIN}(b) shows the base design cells in which the material distribution is manipulated to solve the optimization problem. The content of each base cell is duplicated throughout $\Omega_{\text{dR,1}}$ and $\Omega_{\text{dR,2}}$ to construct the material distribution (topological phases) used when solving \eqref{EQN:HELMHOLTZ}. For the example treated in this letters C3v-symmetry is imposed on both base cells. The designable region is colored grey and the mirror symmetries are shown using dashed lines. An example of a design for one phase and its symmetry is illustrated.} 

\noindent The optimization problem is written as,

\begin{eqnarray}
	\max_{\xi(\textbf{r}) \in [0,1]} \ \ \ &\Phi_{\text{Total}}(\xi)& = \sum_{i=1}^{3} \Phi_{\text{Max}_i}(\xi) - \Phi_{\text{Min}}(\xi), \label{EQN:OPTIMIZATION_PROBLEM} \\  
	\text{s.t.} \ \ \ &\Phi_{\text{BG}}(\xi)& \leq \gamma_1, \label{EQN:OPTIMIZATION_PROBLEM_CONSTRAINT_1} \\  
	&\gamma_2& < \Phi_{\text{Max}_1}(\xi) / \Phi_{\text{Max}_2}(\xi) < \gamma_3, \label{EQN:OPTIMIZATION_PROBLEM_CONSTRAINT_2}
\end{eqnarray}

\noindent where $\Phi_{\text{Total}}$ is the objective function consisting of a linear combination of the terms $\Phi_{\text{Max}_i}, i \in \lbrace 1,2,3 \rbrace$ and $\Phi_{\text{Min}}$, all of which are integrals of the field intensity magnitude over $\Omega_{\text{Max}_i}, i \in \lbrace 1,2,3 \rbrace$ and $\Omega_{\text{Min}}$, while $\Phi_{\text{BG}}$ denotes the integral of the field intensity magnitude over $\Omega_{\text{BG}}$, see Fig. \ref{FIG:MODEL_TARGET_DESIGN_DOMAIN}\textcolor{black}{(d)}. The constants $\gamma_j > 0, j \in \lbrace 1,2,3 \rbrace$ control the constraints \eqref{EQN:OPTIMIZATION_PROBLEM_CONSTRAINT_1}-\eqref{EQN:OPTIMIZATION_PROBLEM_CONSTRAINT_2} and $\Phi_{\star}$ is calculated as,

\begin{eqnarray} \label{EQN:INTENSITY_INTEGRAL}
\Phi_{\star}(\xi) &=& \left. \tau_{\star} \int_{\Omega_{\star}} \vert \text{I}(\Psi(\xi)) \vert \text{d}\textbf{r} \middle/ \int_{\Omega_{\star}} \text{d}\textbf{r} \right. , \\ 
\star &\in& \lbrace {\text{Max}_i},{\text{Min}},{\text{BG}} \rbrace, i \in \lbrace 1,2,3 \rbrace. \nonumber
\end{eqnarray}

\noindent Here $\text{I}(\Psi(\xi))$ denotes the field intensity and $\tau_{\star}$ a set of scaling constants. 
The choice of $\Phi_{\text{Total}}$ leads to a maximization of the energy transmitted into $\Omega_{\text{Max}_1}$ and $\Omega_{\text{Max}_2}$ along with a simultaneous minimization of the energy transmitted into $\Omega_{\text{Min}}$. That is, in order to maximize $\Phi_{\text{Total}}$ any field emitted by $\textbf{P}_{\text{S}}$, propagating along the interface between $\Omega_{\text{dR},1}$ and $\Omega_{\text{dR},2}$, must keep $\Omega_{\text{dR},1}$ on its right hand side and $\Omega_{\text{dR},2}$ on its left hand side at all times. This prohibits a change in propagation direction, which would occur if the field propagated along the interface between $\Omega_{\text{dR},1}$ and $\Omega_{\text{dR},2}$ to $\Omega_{\text{Min}}$, as the spatial symmetry is inverted at the centre of the material slab. This in turn promotes back-scattering protected transport of energy along the interface. The constraint \eqref{EQN:OPTIMIZATION_PROBLEM_CONSTRAINT_1} ensures that a bulk-bandgap exists in both topological phases, as energy is prohibited from propagating into $\Omega_{\text{BG}}$. The constraint \eqref{EQN:OPTIMIZATION_PROBLEM_CONSTRAINT_2} may be used to control the ratio of the intensity transmitted to $\Phi_{\text{Max}_1}$ and $\Phi_{\text{Max}_2}$, respectively. 

\textcolor{black}{The design problem, \eqref{EQN:HELMHOLTZ}-\eqref{EQN:INTENSITY_INTEGRAL}, is implemented and solved in COMSOL Multiphysics 5.3 using the deterministic gradient-based optimization method, the globally convergent method of moving asymptotes (GCMMA) \cite{SVANBERG_2002} to solve \eqref{EQN:OPTIMIZATION_PROBLEM}-\eqref{EQN:OPTIMIZATION_PROBLEM_CONSTRAINT_2}. The objective function gradients are calculated efficiently using adjoint sensitivity analysis \cite{JENSEN_SIGMUND_2011}. A physically admissible final design, consisting solely of solid and air and free of numerical artefacts, is assured using the projection and filtering procedure outlined in \cite{CHRISTIANSEN_SIGMUND_2016,WANG_ET_AL_2011,GUEST_ET_AL_2004}.}

\begin{figure}[h]
	\centering
	{
		\includegraphics[width=0.47\textwidth]{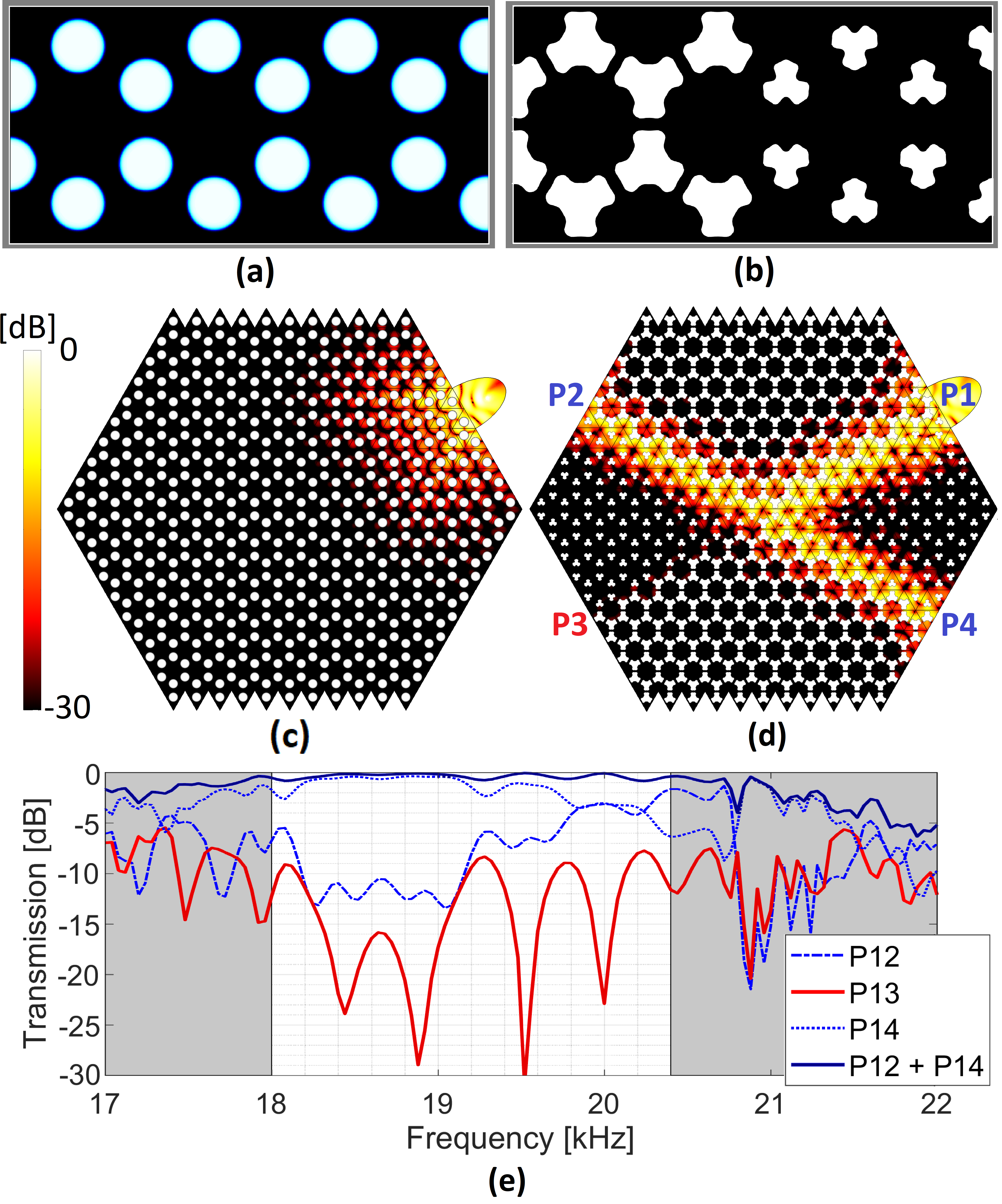} 
		\caption{\textbf{(a)} Initial and \textbf{(b)} optimized material configuration in a super cell consisting of the two topological phases [black] air, [white] solid, [shades of blue] air and solid mixture. \textbf{(c-d)} Sound pressure at $f_0 = 20$ kHz for the \textbf{(c)} initial and \textbf{(d)} optimized material distribution in $\Omega_{\text{dR},1} \bigcup \Omega_{\text{dR},2}$. [colormap] Pressure magnitude, [white] solid material. \textbf{(d)} The targeted backscattering protected edge-state energy transport is observed. \textbf{(e)} Transmission to ports 2, 3 and 4, as a function of frequency. normalized to the energy flowing through port 1 (P1). Port numbering is shown in \textbf{(d)}. \label{FIG:HEXAGONAL_CONFIGURATION_PERFORMANCE}}
	}
\end{figure} 

\begin{figure}[h]
	\centering
	{
		\includegraphics[width=0.48\textwidth]{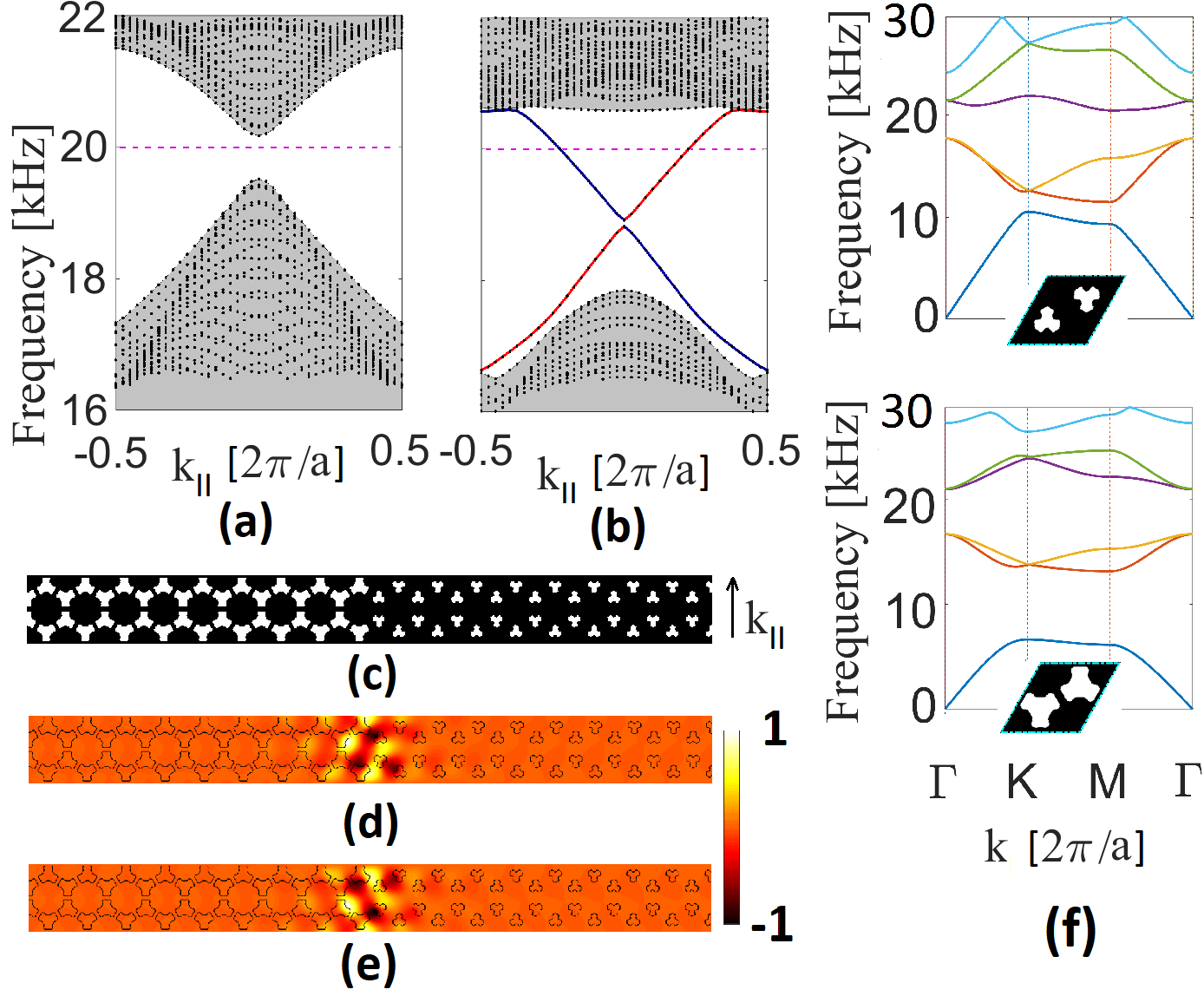}
		\caption{Supercell band structure for the \textbf{(a)} initial material configuration and \textbf{(b)} optimized TI design. The edge-state bands are colored indicating [red] positive and [blue] negative pseudo spin-1/2. [white] bulk-bandgap. \textbf{(c)} Supercell used to compute \textbf{(b)}. \textbf{(d-e)} Normalized eigenmodes at $f_0 =$ 20 kHz. \textcolor{black}{\textbf{(f)} Band structures for the two crystal phases.} \label{FIG:BANDSTRUCTURE_EIGENMODES}}
	}
\end{figure}

For the TI considered in the following \eqref{EQN:HELMHOLTZ}-\eqref{EQN:INTENSITY_INTEGRAL} is solved with $\lbrace a = 0.01 \text{ m}, f_0 = 20 \text{ kHz}, c = 343 \text{ m}/\text{s}, \alpha = 6 \text{ dB}/\lambda \ \forall \ \textbf{r} \in \Omega_{\text{A}}, \alpha = 0 \text{ dB}/\lambda \ \forall \ \textbf{r} \in \Omega_{\text{dR,1}} \bigcup \Omega_{\text{dR,2}}, \gamma_1 = 0.04, \gamma_2 = 0.3, \gamma_3 = 1.7, \tau_{\text{Max}_1} = 1, \tau_{\text{Max}_2} = 1, \tau_{\text{Max}_3} = 0.1, \tau_{\text{Min}} = 4, \tau_{\text{BG}} = 1 \rbrace$. The initial $\xi(\textbf{r})$-layout, shown in Fig. \ref{FIG:HEXAGONAL_CONFIGURATION_PERFORMANCE}(a), is chosen to constitute a crystal with a bulk-bandgap at $f_0$ [see the band structure in Fig. \ref{FIG:BANDSTRUCTURE_EIGENMODES}(a)]. The final material layout obtained from the optimization process is shown in Fig. \ref{FIG:HEXAGONAL_CONFIGURATION_PERFORMANCE}(b), with white/black representing solid/air.

The max-normalized pressure field at $f_0 = 20$ kHz, along with the initial and optimized material configurations in $\Omega_{\text{dR},1} \bigcup \Omega_{\text{dR},2}$ are shown in Figs. \ref{FIG:HEXAGONAL_CONFIGURATION_PERFORMANCE}(c) and \ref{FIG:HEXAGONAL_CONFIGURATION_PERFORMANCE}(d), respectively. The bulk-bandgap of the initial material configuration is clearly observed. For the optimized TI design it is clear that the vast majority of the energy flowing into port 1 (P1) is transmitted to either port 2 (P2) or port 4 (P4). Simultaneously a bulk-bandgap is observed for both phases of the TI. Figure \ref{FIG:HEXAGONAL_CONFIGURATION_PERFORMANCE}(e) presents a frequency sweep of the transmission to ports 2, 3 and 4, normalized to the power flowing through port 1: $10\log_{10}(|\text{P}_x|/|\text{P}_1|)$. It is seen that 99.5\% of the acoustic power is transmitted from port 1 to ports 2 and 4 at $f_0 = 20$ kHz. Further, it is seen that the transmission does not drop below $-0.85$ dB from $18$ kHz to $20.4$ kHz. 

The above discussion demonstrates that the top-down approach results in the desired macroscopic response. However, as no explicit requirements on the existence of TI effects were included in the optimization formulation, the macroscopic response could in principle be based on other effects. That a TI has indeed appeared spontaneously through the optimization process is revealed in the following analysis.

Figure \ref{FIG:BANDSTRUCTURE_EIGENMODES}(b) shows the band structure diagram, calculated for the TI super cell shown in Fig. \ref{FIG:BANDSTRUCTURE_EIGENMODES}(c) using periodic boundary conditions on the top and bottom edges and Neumann conditions on left and right edges. The bulk-band regions are colored grey and the two "crossing" symmetry inverted edge-state bands are colored red and blue corresponding to the positive and negative pseudo spin-1/2 edge-state modes, shown for $f_0 = 20 \text{ kHz}$ in Figs. \ref{FIG:BANDSTRUCTURE_EIGENMODES}(d) and \ref{FIG:BANDSTRUCTURE_EIGENMODES}(e), respectively. From Fig. \ref{FIG:BANDSTRUCTURE_EIGENMODES}(b) the bulk-bandgap is seen to stretch from $\approx$18 kHz to $\approx$20.4 kHz, a bandgap of $\approx$$12.5\%$. A narrow gap is seen in the two edge-state bands at the $\text{k}_{\vert\vert} = 0$ point. A similar gap was reported in \cite{HE_2016} where it was explained to originate from the imperfect cladding layer rather than the TI itself. \textcolor{black}{Figure \ref{FIG:BANDSTRUCTURE_EIGENMODES}(f) presents the band structure for the first six bands of each of the two crystal phases constituting the TI, revealing degeneracies for bands 2 and 3 and bands 4 and 5 at the $\Gamma$-point for both crystal phases.}

\begin{figure}[h]
	\centering
	{
		\includegraphics[width=0.47\textwidth]{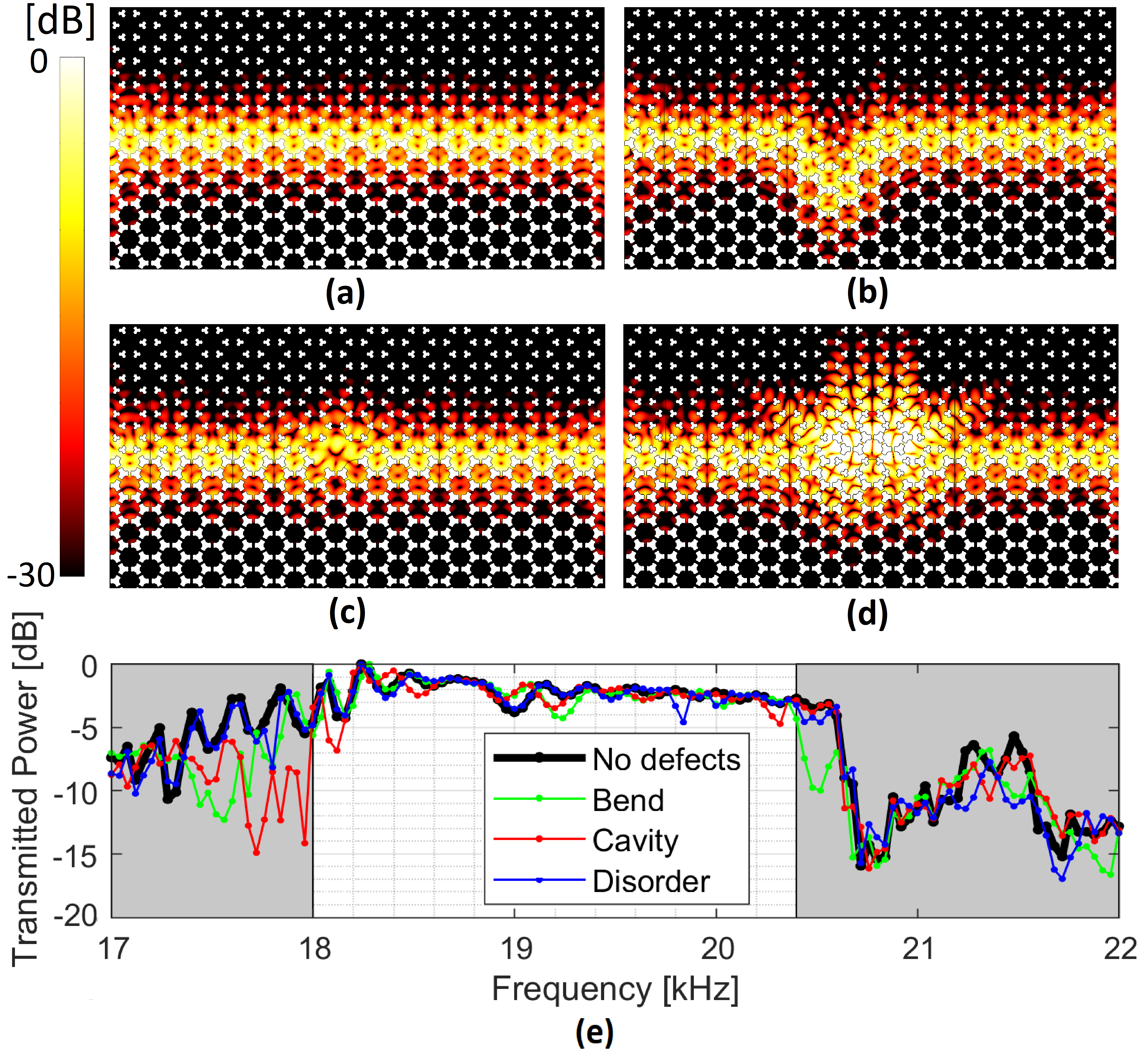}
		\caption{Investigation of robust energy transport in TI edge-state under geometric defects. \textbf{(a-d)} [colormap] Max-normalized sound pressure at 20 kHz, [white] solid material. \textbf{(a)} Unperturbed TI. \textbf{(b)} TI with bend. \textbf{(c)} TI with cavity. \textbf{(d)} TI with disorder. \textbf{(e)} Max-normalized transmitted power, recorded after the material slab, for the configurations shown in \textbf{(a-d)}.\label{FIG:SIMPLE_PERTURBED_CONFIGURATION_PERFORMANCE} }
	}
\end{figure} 

To further investigate if a TI supporting geometrically robust backscattering protected transport of acoustic energy has been designed, a series of studies on the effect of introducing defects in the TI are performed. Figure \ref{FIG:SIMPLE_PERTURBED_CONFIGURATION_PERFORMANCE} presents four examples, with Fig. \ref{FIG:SIMPLE_PERTURBED_CONFIGURATION_PERFORMANCE}(a) showing a slab of the TI without any defects as a reference, while Figs. \ref{FIG:SIMPLE_PERTURBED_CONFIGURATION_PERFORMANCE}(b)-\ref{FIG:SIMPLE_PERTURBED_CONFIGURATION_PERFORMANCE}(d) show a bend, cavity and disorder defect, respectively. \textcolor{black}{The three defects all preserve the symmetry of the bulk materials and are shown in Fig. \ref{FIG:MODEL_TARGET_DESIGN_DOMAIN}(a), where they are highlighted using green.} The slabs are excited by a point source positioned 0.3a from their left edge, at the interface of the two topological phases. The power, transmitted through the TI, is computed at the right side of the slab for each configuration. The results of these computations are reported in Fig. \ref{FIG:SIMPLE_PERTURBED_CONFIGURATION_PERFORMANCE}(e), max-normalized with respect to the non-defect TI. 

Figure \ref{FIG:SIMPLE_PERTURBED_CONFIGURATION_PERFORMANCE}(e), reveals good agreement of the transmitted power inside the bulk-bandgap across the four cases. The largest deviation between the non-defect and defect structures is 2.5 dB, and intervals showing less than 0.25 dB deviation are observed. These results support that a TI offering backscattering protected propagation has been designed. The differences in transmission seen in Fig. \ref{FIG:SIMPLE_PERTURBED_CONFIGURATION_PERFORMANCE}(e) are orders of magnitude smaller than the differences observed across the majority of the band of operation for similar defects in a traditional phononic crystal wave-guide, with a worst case value of more than 25 dB reported in \cite{HE_2016}. 

\begin{figure}[H]
	\centering
	{
		\includegraphics[width=0.43\textwidth]{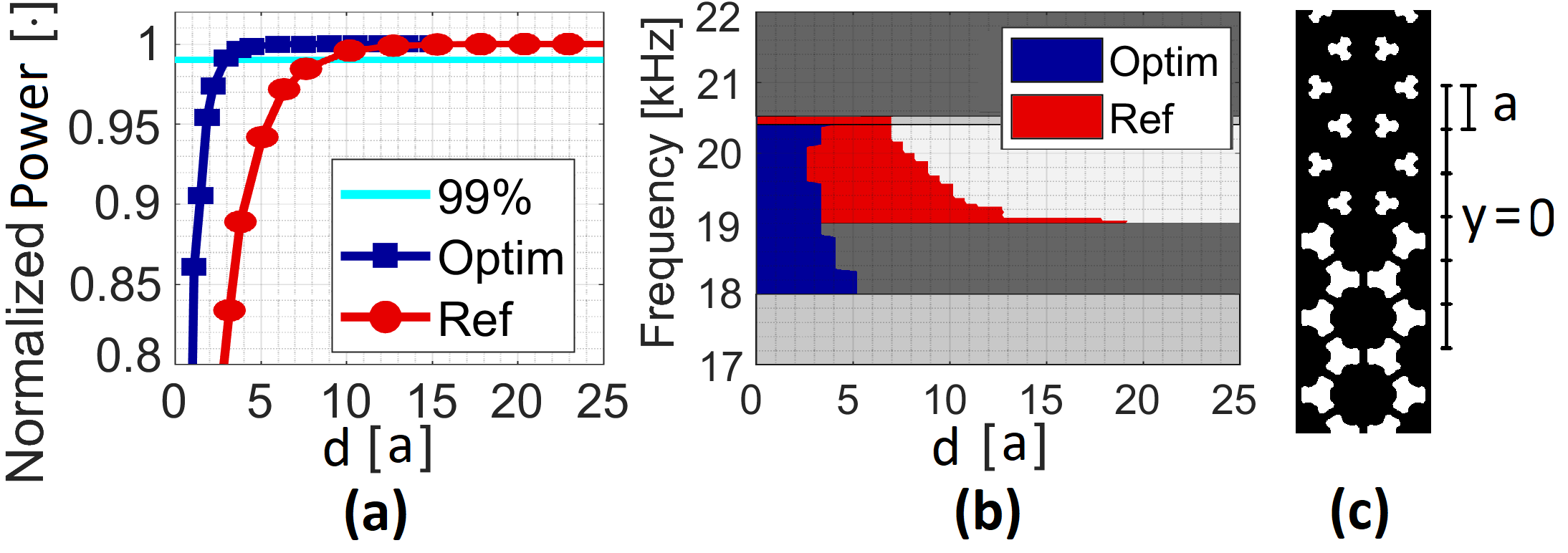}
		\caption{Spatial field confinement of the TI edge-state. \textbf{(a)} Fraction of the total power contained within a distance $d$ from the TI interface edge, $\text{P}_{\text{f}}(d)$, at $f_0 = 20$ kHz, for [Optim] the proposed TI design and [Ref] the TI in \cite{HE_2016}. Map of the distance $d$ versus frequency, within which more than $99\%$ of the total power is confined, $\text{P}_{\text{f}}(d)>0.99$ for [Optim] the optimized TI design and [Ref] the reference. \textbf{(c)} TI super cell, [white] solid material, [black] air.\label{FIG:FIELD_CONFINEMENT}}
	}
\end{figure} 

An important aspect to consider when designing systems for energy/information transport, such as wave-guides, is the footprint of the system. In the present context, the footprint refers to how wide the material slab must be to confine a certain fraction of the transported energy. From Fig. \ref{FIG:SIMPLE_PERTURBED_CONFIGURATION_PERFORMANCE}(a), the pressure field in the TI appears to be confined (to a 30 dB level) inside approximately $3a$ from the TI interface edge. An investigation of the spatial confinement of the field is performed using the TI from \cite{HE_2016} as a reference. This is done by calculating the fraction of the total power flowing through the TI within a distance $d$ from the TI interface edge [see the illustration in Fig. \ref{FIG:FIELD_CONFINEMENT}(c)],

\begin{equation} \label{EQN:FIELD_CONFINEMENT}
\text{P}_{\text{f}}(d) = \left. \int_{-d}^{d} \textbf{n} \cdot \textbf{I}(y) \ \text{d}y \middle/ \int_{-\infty}^{\infty} \textbf{n} \cdot \textbf{I}(y) \ \text{d}y \right. , \ \ d>0.
\end{equation}

The results for $f_0 = 20$ kHz are shown in Fig. \ref{FIG:FIELD_CONFINEMENT}(a). From here it is observed that more than 99\% of the power is contained within $d \approx 3a$ for the TI design proposed in this letter, while for the reference $d \approx 9a$ is required. A map showing the distance $d$ within which 99\% or more of the power is confined versus frequency for both TIs is provided in Fig. \ref{FIG:FIELD_CONFINEMENT}(b). From this map it is seen that at most a distance of $6a$ is required to contain 99\% of the power for the proposed TI.
	
In summary, this letter reports on the design of a topological insulator using a top-down approach based on density based topology optimization. The approach directly targets the desired effect of backscattering protected, directional energy transport. That the effect is achieved by the resulting TI is demonstrated through numerical studies. Experimental validation of the approach may be found in \cite{CHRISTIANSEN_SIGMUND_2016b} where a meta-material exhibiting negative refraction is considered. 

The proposed design approach is trivially extendible to photonics, assuming TE or TM polarized light. Further, by introducing additional design constraints and goals it is straightforwardly extendable to e.g. consider global defects in the TI or to target a maximization of the operational bandwidth in the design process. Hence, the approach has freedom to tailor TIs to operate under alternative conditions.

\begin{acknowledgments}
The authors acknowledge discussions with S.
Stobbe and support from NATEC (NAnophotonics for Terabit Communications) Centre (Grant No. 8692).
\end{acknowledgments}

\bibliographystyle{apsrev4-1} 
\bibliography{References}   

\end{document}